\documentclass[10pt,aps,prb,reprint,superscriptaddress,floatfix]{revtex4-1}

\usepackage[UKenglish]{babel}
\usepackage[utf8]{inputenc}
\usepackage[T1]{fontenc}
\usepackage{graphicx}
\usepackage{amssymb}
\usepackage{amsmath}
\usepackage{hyperref}
\usepackage{enumitem}

\newif\ifuseexttikz

\useexttikztrue

\ifuseexttikz
\else
\usepackage{pgfplots}
\usepackage{tikz}
\usetikzlibrary{external} 
\fi

\newcommand{\vcenteredinclude}[1]{\begingroup
\setbox0=\hbox{#1}%
\parbox{\wd0}{\box0}\endgroup}

\hypersetup{pdfauthor={C. Hubig, I. P. McCulloch, U. Schollwoeck},pdftitle={Generic Construction of Efficient Matrix Product Operators}}

\usepackage{enumitem}
\usepackage{color}
\usepackage[amssymb]{SIunits}
\newcommand{\tikzpic}[2]{
  \ifuseexttikz
  \vcenteredinclude{\includegraphics{#1}}
  \else
  \tikzsetnextfilename{#1}
  #2
  \fi
}

\begin{document}
\title{Generic Construction of Efficient Matrix Product Operators}
\author{C.\ Hubig}
\email{c.hubig@physik.uni-muenchen.de}
\affiliation{Department of Physics and Arnold Sommerfeld Center for
  Theoretical Physics, Ludwig-Maximilians-Universit\"at M\"unchen,
  Theresienstrasse 37, 80333 M\"unchen, Germany}
\author{I.\ P.\ McCulloch}
\affiliation{Centre for Engineered Quantum Systems, School of Physical
  Sciences, The University of Queensland, Brisbane, Queensland 4072,
  Australia}
\author{U.\ Schollw\"ock}
\affiliation{Department of Physics and Arnold Sommerfeld Center for
  Theoretical Physics, Ludwig-Maximilians-Universit\"at M\"unchen,
  Theresienstrasse 37, 80333 M\"unchen, Germany}
\begin{abstract}
  Matrix Product Operators (MPOs) are at the heart of the
  second-generation Density Matrix Renormalisation Group (DMRG)
  algorithm formulated in Matrix Product State language. We first
  summarise the widely known facts on MPO arithmetic and
  representations of single-site operators. Second, we introduce three
  compression methods (Rescaled SVD, Deparallelisation and
  Delinearisation) for MPOs and show that it is possible to construct
  efficient representations of arbitrary operators using MPO
  arithmetic and compression. As examples, we construct powers of a
  short-ranged spin-chain Hamiltonian, a complicated Hamiltonian of a
  two-dimensional system and, as proof of principle, the long-range
  four-body Hamiltonian from quantum chemistry.
\end{abstract}
\date{\today}
\maketitle

\section{\label{sec:intro}Introduction}

Since its introduction in 1992, the Density Matrix Renormalisation
Group\cite{white92:_densit} (DMRG) algorithm has been extremely
successful at the solution of one-dimensional quantum mechanical
problems.\cite{schollwoeck05} Following the
connection\cite{oestlund95:_therm_limit_densit_matrix_renor} between
the original DMRG algorithm and the variational class of Matrix
Product States (MPS), a series of second-generation DMRG algorithms
has been developed\cite{mcculloch07:_from, pirvu10:_matrix,
  schollwoeck11, white05:_densit,
  dolfi12:_multig_algor_tensor_networ_states, stoudenmire13:_real,
  schollwoeck11} which explicitly build on the underlying tensor
structure. In these second-generation algorithms, both the current
variational state as well as the Hamiltonian operator are represented
as tensor networks, namely MPS and Matrix Product Operators (MPO).

As such, the correct construction of the MPO representation of the
Hamiltonian at hand is the starting point of any DMRG
calculation. This construction can be done fairly easily by hand for
short-range Hamiltonians, if necessary with the help of a finite-state
machine\cite{schollwoeck11, froewis10:_tensor, crosswhite08:_apply}
which generates the required terms in the MPO. However, these
finite-state machines can very quickly become extremely complicated
(see e.g.\ Ref.~\onlinecite{motruk16:_densit} Fig.~7, 9 and 10 for
automata to generate interactions on a two-dimensional
cylinder). Other analytical approaches\cite{keller15:_hamil,
  kin-lic16:_matrix_produc_operat_matrix_produc} to construct the MPO
representation of in particular quantum chemistry Hamiltonians require
individual treatment of each system and type of interaction by hand.

In this paper, we will present a generic method to construct arbitrary
MPOs based solely on a) the definition of appropriate single-site
operators (such as $c^\dagger_i$ or $s^z_i$) and b) the implementation
of a model-independent MPO arithmetic. We will show that using these
two ingredients, it is possible to efficiently construct the optimal
representations of small powers of one-dimensional Hamiltonians and of
medium-range Hamiltonians on two-dimensional cylinders. We further
provide a proof-of-principle that the constructive approach is also
able to generate the optimal representation for the four-body quantum
chemistry Hamiltonian with long-range interactions.

The outline of the paper is as follows: In Section \ref{sec:mpodef} we
define MPOs as widely used in the literature. Sections \ref{sec:sso}
and \ref{sec:arithmetic} summarise and supplement the existing works
on the construction of fundamental single-site operators such as
$c_i^\dagger$ in MPO form as well as the addition and multiplication
of arbitrary MPOs. After such an addition or multiplication,
compression using one of the three compression methods specifically
adapted to MPOs as laid out in Section \ref{sec:trunc} brings the
operator representation back into its most efficient form. We give
examples of the resulting MPOs in Section \ref{sec:example} for a
spin-chain with nearest-neighbor interactions, the Fermi-Hubbard model
on a cylinder in hybrid real- and momentum space and the full quantum
chemistry Hamiltonian. Section \ref{sec:variance} details an algorithm
to reduce numerical errors while calculating the variance
$\langle O^2 \rangle - \langle O \rangle^2$ of a MPO -- of particular
interest here is the Hamiltonian $\hat H$ represented as a
MPO. Finally, we conclude in Section \ref{sec:conclusions}.

\section{\label{sec:mpodef}Matrix Product Operators (MPO)}

For a detailed introduction to the Density Matrix Renormalization
Group (DMRG) and in particular the second-generation algorithms based
on Matrix Product States (MPS) and Matrix Product Operators (MPOs), we
refer to an existing review\cite{schollwoeck11} as well as a
DMRG-centered overview of the
implementation.\cite{hubig15:_stric_dmrg} Here, we will only define
the basic structure of matrix product operators.

\begin{figure}
  \centering
  \tikzpic{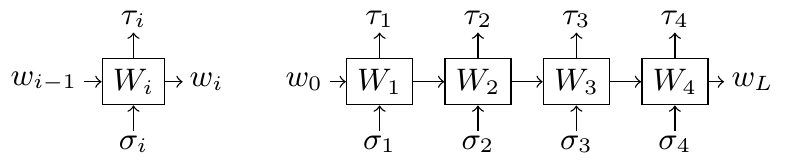}{
    \begin{tikzpicture}[baseline=-2]
      \node (wi) [draw] at (0,0){$W_i$};
      \draw [<-] (wi) -- node[near end, left]{$w_{i-1}$}(-0.5,0);
      \draw [->] (wi) -- node[near end, right]{$w_i$}(0.5,0);
      \draw [->] (wi) -- node[near end, above]{$\tau_i$}(0,0.5);
      \draw [<-] (wi) -- node[near end, below]{$\sigma_i$}(0,-0.5);

      \node (w1) [draw] at (2.5,0){$W_1$};
      \node (w2) [draw] at (3.5,0){$W_2$};
      \node (w3) [draw] at (4.5,0){$W_3$};
      \node (w4) [draw] at (5.5,0){$W_4$};

      \draw [->] (w1) -- node[near end, above]{$\tau_1$} +(0,0.5);
      \draw [<-] (w1) -- node[near end, below]{$\sigma_1$} +(0,-0.5);
      \draw [->] (w2) -- node[near end, above]{$\tau_2$} +(0,0.5);
      \draw [<-] (w2) -- node[near end, below]{$\sigma_2$} +(0,-0.5);
      \draw [->] (w3) -- node[near end, above]{$\tau_3$} +(0,0.5);
      \draw [<-] (w3) -- node[near end, below]{$\sigma_3$} +(0,-0.5);
      \draw [->] (w4) -- node[near end, above]{$\tau_4$} +(0,0.5);
      \draw [<-] (w4) -- node[near end, below]{$\sigma_4$} +(0,-0.5);

      \draw [<-] (w1) -- node[near end, left]{$w_0$} +(-0.5,0);
      \draw [->] (w1) -- (w2);
      \draw [->] (w2) -- (w3);
      \draw [->] (w3) -- (w4);
      \draw [->] (w4) -- node[near end, right]{$w_L$} +(0.5,0);
    \end{tikzpicture}
  }
  \caption{Left: Graphical representation of one component tensor
    $W_i$ of a MPO. Right: Contraction corresponding to the
    matrix-matrix products in Eq.~\eqref{eq:mpodef:wwwc} of multiple
    component tensors. Labels on the legs denote the basis on this
    leg. Arrows indicate whether the leg is incoming or outgoing and
    are largely only relevant when implementing quantum number
    conservation. In this convention, MPOs act on Matrix Product
    States from above, with the latter having outgoing physical
    indices.}
  \label{fig:mpo}
\end{figure}
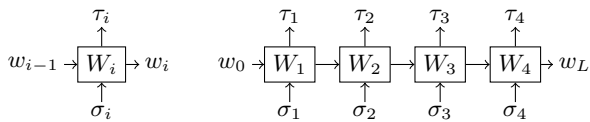

Given a set of $L$ local Hilbert spaces $\mathcal{H}_i, i \in [1, L]$
of dimension $d_i$ each and an operator $\hat H$ which acts on the
tensor product space $\mathcal{H} = \otimes_i \mathcal{H}_i$, we can
write the operator $\hat H$ as

\begin{equation}
  \hat H = \sum_{\boldsymbol{\sigma \tau}} c^{\boldsymbol{\sigma \tau}} |\boldsymbol{\tau}\rangle\langle\boldsymbol{\sigma}|
\end{equation}

where $|\boldsymbol{\tau}\rangle$ enumerates the (product) basis
states of $\mathcal{H}$ and $\langle\boldsymbol{\sigma}|$ enumerates
the basis states of the dual space of $\mathcal{H}$. We can decompose
each basis vector $|\boldsymbol{\tau}\rangle$ as the tensor product of
basis vectors on the individual local spaces as

\begin{equation}
  |\boldsymbol{\tau}\rangle = \bigotimes_{i=1}^L |\tau_i\rangle = |\tau_1 \ldots \tau_L \rangle
\end{equation}

which leads to

\begin{equation}
  \hat H = \sum_{\sigma_1\tau_1} \cdots \sum_{\sigma_L \tau_L} c^{\sigma_1 \ldots \sigma_L}_{\tau_1 \ldots \tau_L} |\tau_1 \ldots \tau_L \rangle \langle \sigma_1 \ldots \sigma_L |.
\end{equation}

This form is still entirely generic.\cite{schollwoeck05} The
coefficient $c \in \mathbb{C}^{\prod_i d_i^2}$ may now be decomposed
as a set of \emph{matrix products}. That is, on each site $i$ and for
every combination of local states
$\{|\tau_i\rangle, \langle\sigma_i|\}$, we introduce a set of matrices
$\left(W_i^{\sigma_i \tau_i}\right)_{w_{i-1},w_i}$ with the property
that their matrix-matrix product equals a specific element of the $c$
tensor:

\begin{align}
  \sum_{\boldsymbol{w}} \left(W_1^{\sigma_1 \tau_1}\right)_{w_0,w_1} \cdot & \left(W_2^{\sigma_2 \tau_2}\right)_{w_1, w_2} \cdots \left(W_L^{\sigma_L \tau_L}\right)_{w_{L-1},w_L}  \nonumber \\
  & = c^{\sigma_1 \ldots \sigma_L}_{\tau_1 \ldots \tau_L} \quad. \label{eq:mpodef:wwwc}
\end{align}

The tensor $W_i$ is then a rank-4 tensor with two \emph{physical
  indices} $\sigma_i$ and $\tau_i$ while the two matrix indices above
are now called \emph{MPO bond indices} and will be labelled $w_{i-1}$
and $w_{i}$. In order for the above product of matrices to result in a
scalar value for a given set of $\boldsymbol{\tau}$ and
$\boldsymbol{\sigma}$, we need $w_0$ and $w_L$ to be 1-dimensional
dummy indices. Each tensor $W_i$ can be represented graphically by a
square with four legs, cf.\ Fig.~\ref{fig:mpo}. Connecting legs of two
tensors corresponds to a tensor contraction over the associated
indices.

The relevant insight is that for a large class of operators, including
all Hamiltonians with short-range interactions in one dimension, the
required \emph{MPO bond dimension}, i.e.\ the size of matrices
$W_i^{\sigma_i \tau_i}$ in Eq.~\eqref{eq:mpodef:wwwc}, to reproduce
the original tensor $c$, is both small ($\approx 5$) and constant in
the size $L$ of the system. For long-range interactions, the size of
the matrices usually only grows polynomially in the range of the
interaction.\footnote{Some specific long-range interaction terms can be
  encoded efficiently in constant space, but this is not generally the
  case} Constructing the set of tensors $W_i$ which faithfully
reproduce the desired operator $\hat H$ at minimal MPO bond dimensions
$w_i$ will be discussed in this paper.

\section{\label{sec:sso}Construction of Single-Site Operators}
\begin{figure}
  \centering
  \tikzpic{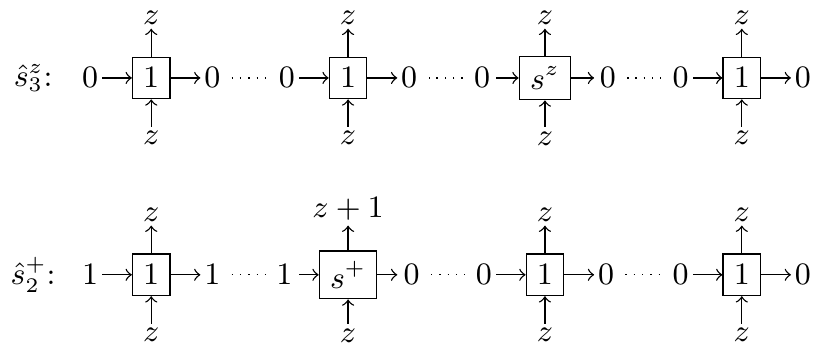}{
    \begin{tikzpicture}[baseline=-2]
      \node at (-1.2,0) {$\hat s^z_3$:};
      \node (i1) [draw] at (0,0){$1$};
      \draw [<-] (i1) -- node[near end, left]{0}(-0.5,0);
      \draw [->] (i1) -- node[near end, right](i1r){0}(0.5,0);
      \draw [->] (i1) -- node[near end, above]{$z$}(0,0.5);
      \draw [<-] (i1) -- node[near end, below]{$z$}(0,-0.5);

      \node (i2) [draw] at (2,0){$1$};
      \draw [<-] (i2) -- node[near end, left](i2l){0}(1.5,0);
      \draw [->] (i2) -- node[near end, right](i2r){0}(2.5,0);
      \draw [->] (i2) -- node[near end, above]{$z$}(2,0.5);
      \draw [<-] (i2) -- node[near end, below]{$z$}(2,-0.5);

      \draw [dotted] (i1r) -- (i2l);

      \node (z3) [draw] at (4,0){$s^z$};
      \draw [<-] (z3) -- node[near end, left](z3l){0}(3.5,0);
      \draw [->] (z3) -- node[near end, right](z3r){0}(4.5,0);
      \draw [->] (z3) -- node[near end, above]{$z$}(4,0.5);
      \draw [<-] (z3) -- node[near end, below]{$z$}(4,-0.5);

      \draw [dotted] (z3l) -- (i2r);

      \node (i4) [draw] at (6,0){$1$};
      \draw [<-] (i4) -- node[near end, left](i4l){0}(5.5,0);
      \draw [->] (i4) -- node[near end, right]{0}(6.5,0);
      \draw [->] (i4) -- node[near end, above]{$z$}(6,0.5);
      \draw [<-] (i4) -- node[near end, below]{$z$}(6,-0.5);

      \draw [dotted] (z3r) -- (i4l);

      \node at (-1.2,-2) {$\hat s^+_2$:};
      \node (i1) [draw] at (0,-2){$1$};
      \draw [<-] (i1) -- node[near end, left]{1}(-0.5,-2);
      \draw [->] (i1) -- node[near end, right](i1r){1}(0.5,-2);
      \draw [->] (i1) -- node[near end, above]{$z$}(0,-1.5);
      \draw [<-] (i1) -- node[near end, below]{$z$}(0,-2.5);

      \node (z2) [draw] at (2,-2){$s^+$};
      \draw [<-] (z2) -- node[near end, left](z2l){1}(1.5,-2);
      \draw [->] (z2) -- node[near end, right](z2r){0}(2.5,-2);
      \draw [->] (z2) -- node[near end, above]{$z+1$}(2,-1.5);
      \draw [<-] (z2) -- node[near end, below]{$z$}(2,-2.5);

      \draw [dotted] (i1r) -- (z2l);

      \node (i3) [draw] at (4,-2){$1$};
      \draw [<-] (i3) -- node[near end, left](i3l){0}(3.5,-2);
      \draw [->] (i3) -- node[near end, right](i3r){0}(4.5,-2);
      \draw [->] (i3) -- node[near end, above]{$z$}(4,-1.5);
      \draw [<-] (i3) -- node[near end, below]{$z$}(4,-2.5);

      \draw [dotted] (z2r) -- (i3l);

      \node (i4) [draw] at (6,-2){$1$};
      \draw [<-] (i4) -- node[near end, left](i4l){0}(5.5,-2);
      \draw [->] (i4) -- node[near end, right]{0}(6.5,-2);
      \draw [->] (i4) -- node[near end, above]{$z$}(6,-1.5);
      \draw [<-] (i4) -- node[near end, below]{$z$}(6,-2.5);

      \draw [dotted] (i3r) -- (i4l);
    \end{tikzpicture}}

  \caption{Graphical representation of MPOs for $\hat s^z_3$ and
    $\hat s^+_2$ on a four-site system. Numbers and letters $z$ denote
    the incoming and outgoing $S^z$ quantum numbers on each tensor
    leg. By convention, the leftmost MPO bond index $w_0$ transforms
    the same as the represented operator, while the rightmost MPO bond
    index $w_L$ always transforms as the vacuum of the system. The MPO
    acts on the MPS below it, mapping states with $S^z = z$ to those
    with $S^z = z+1$ in the second example on the second site. Dotted
    lines indicate the contractions which would result in the full
    coefficient tensor $c$ from Eq.~\eqref{eq:mpodef:wwwc}.}
  \label{fig:sso:qnum}
\end{figure}
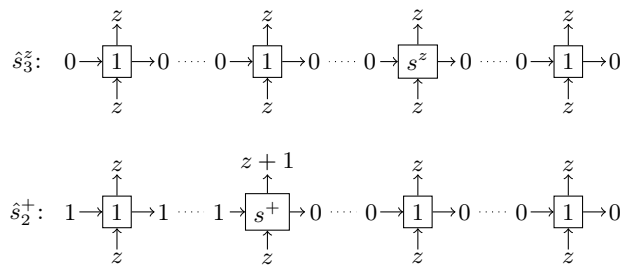
The representation of single-site operators as MPOs is relatively
straightforward in general and mostly already widely known. In this
section we summarise the existing, though not necessarily published,
results in this area.

To construct the MPO representation of a single-site operator, we will
first focus on a homogeneous $S=1$ spin
chain. Subsection~\ref{sec:sso-ferm} contains the transformation of
fermionic operators using a Jordan-Wigner string. In
subsection~\ref{sec:sso-nonh} we explain how to handle non-homogeneous
systems, such as chains of alternating of $S=1$ spins with
$s=\frac{1}{2}$ spins at the ends or mixed fermion-boson systems.

Let us start with the construction of $\hat s^z_i$ for a $S=1$ spin
chain. The first ingredient is the representation of $\hat s^z$ as a
matrix on a local Hilbert space. This is straightforwardly given as
$s^z = \mathrm{diag}(1, 0, -1)$. Secondly, we need the matrix
representation $1_3$ of the identity operator $\hat 1$ on
this local Hilbert space.

For a given fixed $i$, the explicit form of the single-site operator
as $\hat s^z_i = \hat 1_1 \otimes
\hat 1_2 \cdots \hat 1_{i-1} \otimes \hat s^z_i \otimes
\hat 1_{i+1} \cdots \hat 1_L$ (that
is, the identity operator acting on sites $1$ through $i-1$ and $i+1$
through $L$) then corresponds closely to the MPO representation
of $\hat s^z_i$ as

\begin{equation}
  W_{<i} = 1_3 \quad W_i = s^z \quad W_{>i} = 1_3 \quad,
\end{equation}

where the MPO bond indices are 1-dimensional dummy indices and do not
affect the shape of the tensors. The MPO representation of
$\hat s^z_3$ is graphically given in Fig.~\ref{fig:sso:qnum}. For
trivially transforming operators, such as $\hat s^z$ or $\hat n$ which
do not change the quantum numbers of the state, it is entirely
sufficient to store the identity MPO component $1_{d_i}$ and the local
representation (e.g.\ $s^z$) of the operator in question (e.g.\
$\hat s^z$) as rank-4 tensors of size $(1, 1, d_i, d_i)$. One can then
construct the MPO representation on-the-fly.

Operators which do change a quantum number, such as $\hat s^+_i$ or
$\hat c^\dagger_i$, are more complicated. Since each tensor has to
locally preserve symmetries and hence quantum numbers, the additional
quantum number must be carried from the active site $i$ to the left
edge of the system. In turn, the chain of identity operators to the
left of the active site must allow for this quantum number on their
MPO bond indices, while those on the right of the active site only
carry the vacuum quantum numbers (cf.\
Fig.~\ref{fig:sso:qnum}). Therefore, it is necessary to store
different identity operator tensor representations for the left and
right half of the system. It is advisable to simply always store left
and right identities together with the active site tensor, as the
memory requirements of these small tensors are negligible and there is
no need for logic differentiating trivially-transforming and
non-trivially-transforming operators.

Note that the case where no quantum numbers are used (either because
they are not preserved by the system or not supported by the
implementation) is identical to each operator and state transforming
trivially and each leg only carrying a single, appropriately-sized
vacuum sector. In this case, the left and right identity operator
tensor representations are again identical.

\subsection{\label{sec:sso-ferm}Fermionic Operators}
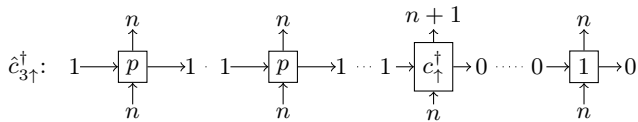
\begin{figure}
  \centering
  \tikzpic{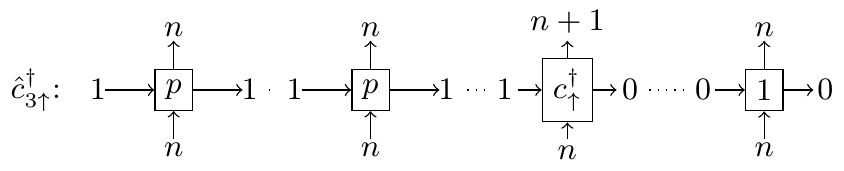}{
    \begin{tikzpicture}[baseline=-2]
      \node at (-1.4,0) {$\hat c^\dagger_{3 \uparrow}$:};
      \node (p1) [draw] at (0,0){$p$};
      \draw [<-] (p1) -- node[near end, left]{1}(-0.7,0);
      \draw [->] (p1) -- node[near end, right](p1r){1}(0.7,0);
      \draw [->] (p1) -- node[near end, above]{$n$}(0,0.5);
      \draw [<-] (p1) -- node[near end, below]{$n$}(0,-0.5);

      \node (p2) [draw] at (2,0){$p$};
      \draw [<-] (p2) -- node[near end, left](p2l){1}(1.3,0);
      \draw [->] (p2) -- node[near end, right](p2r){1}(2.7,0);
      \draw [->] (p2) -- node[near end, above]{$n$}(2,0.5);
      \draw [<-] (p2) -- node[near end, below]{$n$}(2,-0.5);

      \draw [dotted] (p1r) -- (p2l);

      \node (c3) [draw] at (4,0){$c^\dagger_\uparrow$};
      \draw [<-] (c3) -- node[near end, left](c3l){1}(3.5,0);
      \draw [->] (c3) -- node[near end, right](c3r){0}(4.5,0);
      \draw [->] (c3) -- node[near end, above]{$n+1$}(4,0.5);
      \draw [<-] (c3) -- node[near end, below]{$n$}(4,-0.5);

      \draw [dotted] (c3l) -- (p2r);

      \node (i4) [draw] at (6,0){$1$};
      \draw [<-] (i4) -- node[near end, left](i4l){0}(5.5,0);
      \draw [->] (i4) -- node[near end, right]{0}(6.5,0);
      \draw [->] (i4) -- node[near end, above]{$n$}(6,0.5);
      \draw [<-] (i4) -- node[near end, below]{$n$}(6,-0.5);

      \draw [dotted] (c3r) -- (i4l);
    \end{tikzpicture}}

  \caption{Graphical representation of the fermionic creation operator
    $c^\dagger_{3\uparrow}$ on a four-site system. Labels correspond
    to the fermionic particle quantum numbers.}
  \label{fig:sso-ferm:par}
\end{figure}

The implementation of anti-commutation relations for fermionic
operators can also occur at the level of MPO representations of
single-site operators.\cite{dolfi14:_matrix_alps} Proper
anti-commutation \emph{within} the local state space of a single site,
$c^\dagger_{\uparrow, i} c^\dagger_{\downarrow, i} = -
c^\dagger_{\downarrow, i} c^\dagger_{\uparrow, i}$, is contained in
the correct definition of the local site tensor.

Non-local anti-commutation between operators on different sites
requires a defined ordering of all fermionic operators. There is a
natural ordering of operators along the MPO chain from the left to the
right. It then suffices to replace the identities in the previous
section either to the left or to the right of the active site by
parity operators which give a phase of $-1$ if there is an odd number
of fermions on the respective sites.

As an example, consider a product state
$|\psi\rangle = \prod_{i=1}^L \left[ \left(c_{i\uparrow}^\dagger
  \right)^{n_{i\uparrow}} \left( c_{i\downarrow}^\dagger
  \right)^{n_{i\downarrow}} \right] | \mathrm{vacuum} \rangle$.
An operator $c^\dagger_{\downarrow, j}$ applied to $|\psi\rangle$ has
to be commuted past all operators with $i < j$. For each
$n_{i\sigma} \neq 0$, it picks up a minus sign. Each of these signs
can be implemented as the application of the local parity operator
$\hat p_i = (-1)^{n_i}$. The MPO is then constructed as a chain of
parity tensors $p$, the active site tensor $c^\dagger_{\uparrow}$ and
then a chain of right MPO identity components $1_{d_i}$, graphically
represented in Fig.~\ref{fig:sso-ferm:par}. Constructed in such a way,
fermionic MPOs can be treated exactly the same as bosonic MPOs in all
applications that follow.

\subsection{\label{sec:sso-nonh}Non-Homogenous Systems}
\begin{figure}
  \centering
  \tikzpic{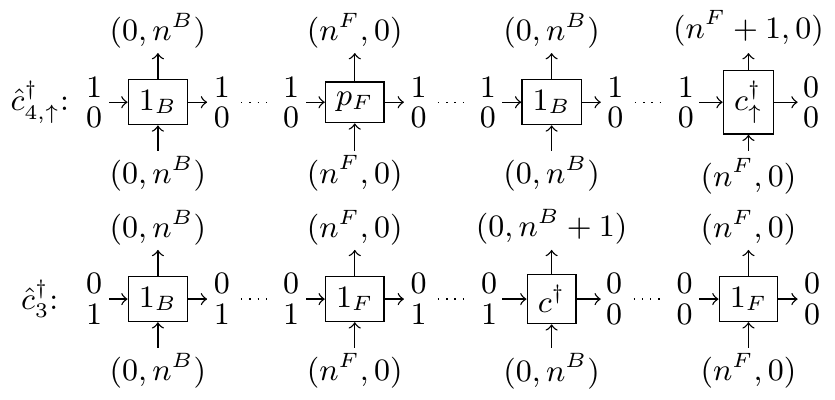}{
    \begin{tikzpicture}[baseline=-2]
      \node at (-1.2,0) {$\hat c^\dagger_{4,\uparrow}$:};
      \node (i1) [draw] at (0,0){$1_B$};
      \draw [<-] (i1) -- node[near end, left]{$\shortstack{1\\0}$}(-0.5,0);
      \draw [->] (i1) -- node[near end, right](i1r){$\shortstack{1\\0}$}(0.5,0);
      \draw [->] (i1) -- node[near end, above]{$(0, n^B)$}(0,0.5);
      \draw [<-] (i1) -- node[near end, below]{$(0, n^B)$}(0,-0.5);

      \node (p2) [draw] at (2,0){$p_F$};
      \draw [<-] (p2) -- node[near end, left](p2l){$\shortstack{1\\0}$}(1.5,0);
      \draw [->] (p2) -- node[near end, right](p2r){$\shortstack{1\\0}$}(2.5,0);
      \draw [->] (p2) -- node[near end, above]{$(n^F, 0)$}(2,0.5);
      \draw [<-] (p2) -- node[near end, below]{$(n^F, 0)$}(2,-0.5);

      \draw [dotted] (i1r) -- (p2l);

      \node (i3) [draw] at (4,0){$1_B$};
      \draw [<-] (i3) -- node[near end, left](i3l){$\shortstack{1\\0}$}(3.5,0);
      \draw [->] (i3) -- node[near end, right](i3r){$\shortstack{1\\0}$}(4.5,0);
      \draw [->] (i3) -- node[near end, above]{$(0, n^B)$}(4,0.5);
      \draw [<-] (i3) -- node[near end, below]{$(0, n^B)$}(4,-0.5);

      \draw [dotted] (p2r) -- (i3l);

      \node (c4) [draw] at (6,0){$c^\dagger_\uparrow$};
      \draw [<-] (c4) -- node[near end, left](c4l){$\shortstack{1\\0}$}(5.5,0);
      \draw [->] (c4) -- node[near end, right](c4r){$\shortstack{0\\0}$}(6.5,0);
      \draw [->] (c4) -- node[near end, above]{$(n^F+1, 0)$}(6,0.5);
      \draw [<-] (c4) -- node[near end, below]{$(n^F, 0)$}(6,-0.5);

      \draw [dotted] (i3r) -- (c4l);

      \node at (-1.2,-2) {$\hat c^\dagger_{3}$:};
      \node (i1) [draw] at (0,-2){$1_B$};
      \draw [<-] (i1) -- node[near end, left]{$\shortstack{0\\1}$} +(-0.5,0);
      \draw [->] (i1) -- node[near end, right](i1r){$\shortstack{0\\1}$} +(0.5,0);
      \draw [->] (i1) -- node[near end, above]{$(0, n^B)$} +(0,0.5);
      \draw [<-] (i1) -- node[near end, below]{$(0, n^B)$} +(0,-0.5);

      \node (i2) [draw] at (2,-2){$1_F$};
      \draw [<-] (i2) -- node[near end, left](i2l){$\shortstack{0\\1}$} +(-0.5,0);
      \draw [->] (i2) -- node[near end, right](i2r){$\shortstack{0\\1}$} +(0.5,0);
      \draw [->] (i2) -- node[near end, above]{$(n^F, 0)$} +(0,0.5);
      \draw [<-] (i2) -- node[near end, below]{$(n^F, 0)$} +(0,-0.5);

      \draw [dotted] (i1r) -- (i2l);

      \node (c3) [draw] at (4,-2){$c^\dagger$};
      \draw [<-] (c3) -- node[near end, left](c3l){$\shortstack{0\\1}$} +(-0.5,0);
      \draw [->] (c3) -- node[near end, right](c3r){$\shortstack{0\\0}$} +(0.5,0);
      \draw [->] (c3) -- node[near end, above]{$(0, n^B+1)$} +(0,0.5);
      \draw [<-] (c3) -- node[near end, below]{$(0, n^B)$} +(0,-0.5);

      \draw [dotted] (i2r) -- (c3l);

      \node (i4) [draw] at (6,-2){$1_F$};
      \draw [<-] (i4) -- node[near end, left](i4l){$\shortstack{0\\0}$} +(-0.5,0);
      \draw [->] (i4) -- node[near end, right]{$\shortstack{0\\0}$} +(0.5,0);
      \draw [->] (i4) -- node[near end, above]{$(n^F, 0)$} +(0,0.5);
      \draw [<-] (i4) -- node[near end, below]{$(n^F, 0)$} +(0,-0.5);

      \draw [dotted] (c3r) -- (i4l);
    \end{tikzpicture}}
  \caption{Graphical representation of $\hat c^\dagger_{4, \uparrow}$
    and $\hat c^\dagger_3$ in a non-homogeneous system. Sites 1 and 3
    may contain bosons, while sites 2 and 4 may contain up to four
    fermions. The creation operator $c^\dagger_{i(\sigma)}$ is taken
    to create either a fermion or boson, depending on the type of the
    site $i$. The parity tensor $p_F = (-1)^{n^F}$ has been used on
    the second site in place of the usual identity to implement
    fermionic anti-commutation rules for $c^\dagger_{4,
      \uparrow}$. Labels denote the $N^F$ and $N^B$ quantum numbers on
    the corresponding indices.}
\label{fig:sso:nonh:ops}
\end{figure}
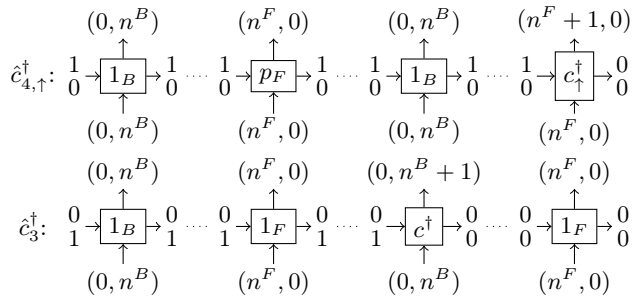

It is possible to simulate non-homogeneous systems using MPS and
MPO. Such a non-homogeneity could be different spin sizes in a spin
chain or the presence of both fermionic and bosonic sites in the
system (the case of non-homogeneous hopping between otherwise
identical sites will be handled later in
Section~\ref{sec:arithmetic}). The former case of non-homogeneity can
be used to represent some experimental systems with alternating $S=1$
and $S=\frac{1}{2}$ spins as well as reduce finite-size effects in
$S=1$ spin chains by placing $S=\frac{1}{2}$ spins at the two
edges. The latter case might be helpful in simulating physical systems
with bosonic and fermionic species, as they commonly occur in
experiments with ultracold atoms.

Suppose we have two types of sites in our system. Even sites may
contain zero, one or two fermions, while odd sites may contain up to a
certain number of bosons.

If we then wish to construct the fermionic creation operator
$c^\dagger_{2i,\uparrow}$, we have to ensure that the identities used
to its left and right match the corresponding physical basis on those
sites. Further, if we use $\mathrm{U}(1)_{N^F} \times
\mathrm{U}(1)_{N^B}$ quantum numbers for fermion and boson number
conservation, the identities to the left need MPO bond indices
transforming as $N^F=1, N^B = 0$. In contrast, if we apply a bosonic
creation operator $c^\dagger_{2i+1}$, the bond indices of those
identities have to transform as $N^F=0, N^B = 1$.\footnote{In this
  specific case, it might be reasonable to set $c^\dagger_{2i+1}$
  equal to zero and call the bosonic creators $a^\dagger$. For spin
  systems, it is however entirely reasonable to have $\hat s^z$ both
  on $S=\frac{1}{2}$ sites (at the edge) and $S=1$ sites (in the bulk)
  of the system.}  Fig.~\ref{fig:sso:nonh:ops} gives examples of those
creation operators.

This has two implications. First, for every type $t^\prime$ of sites
in the system, we need to define an appropriate active tensor
representing (say) $c^\dagger$ acting on a site of this type
$t^\prime$. Second, for every active site type $t^\prime$ on which the
operator acts, we also need to store an appropriate left and right
identity tensor for all types of sites.

Thus, if we have $T$ different types of sites in our system, we need
to store up to $T + 2 T^2$ rank-4 tensors per single-site
operator. However, since these tensors are still only of size $(1, 1,
d_i, d_i)$, and the number of different types $T$ is typically also
small, this is not a concern in practice.

Consider the example of a spin chain with $S=1$ spins in the bulk and
two $S=\frac{1}{2}$ spins at the boundaries. To construct $s^+_i$ on
the fly, we need to store ten rank-4 tensors: First, we need to store
two tensors representing $s^+_i$ acting on sites with $S=1$ and
$S=\frac{1}{2}$. Second, for each of these two, we need to store two
left-identities which we place on sites with $S=1$ and $S=\frac{1}{2}$
respectively to the left of site $i$. Similarly, we need a total of
four right-identities to be placed on sites to the right of site $i$
with $S=1$ and $S=\frac{1}{2}$ respectively for a total of ten tensors
of size $(1, 1, 2S+1, 2S+1)$; in this specific case requiring the
storage of 55 scalar values in total.\footnote{Five of the tensors are
  of size $(1, 1, 2, 2)$ for $S=\frac{1}{2}$ sites, requiring 4
  numbers each. The other five are of size $(1, 1, 3, 3)$ for $S=1$
  sites with 9 scalar entries. We hence need to store $5 \times 4 + 5
  \times 9$ scalar values to represent $s^+_i$ on any site $i$.}

\section{\label{sec:arithmetic}Arithmetic Operations with Matrix Product Operators}
The implementation of arithmetic operations with MPOs is well-known
already\cite{stoudenmire10:_minim, schollwoeck11} and is entirely
independent of the specific form of the operands. In particular, the
implementation can handle single-site operators as constructed in the
previous section and MPOs resulting from earlier arithmetic operations
on equal footing.

\subsection{\label{sec:arithmetic:prod}Products of Matrix Product Operators}
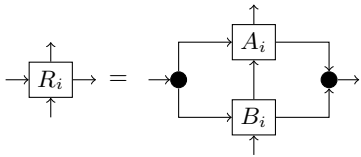
\begin{figure}
  \centering
  \tikzpic{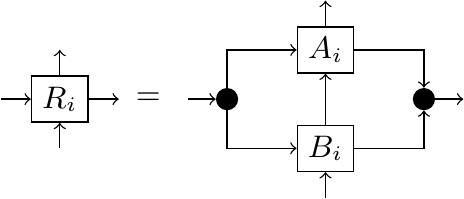}{
    \begin{tikzpicture}[baseline=-2]
      \node (r) [draw] at (-1.7,0){$R_i$};
      \draw [->] (r) -- ++(0,0.5);
      \draw [<-] (r) -- ++(0,-0.5);
      \draw [->] (r) -- ++(0.6,0);
      \draw [<-] (r) -- ++(-0.6,0);
      \node at (-0.8,0) {$=$};
      
      \node (a) [draw] at (1,0.5){$A_i$};
      \node (b) [draw] at (1,-0.5){$B_i$};
      \draw [->] (a) -- ++(0,0.5);
      \draw [<-] (b) -- ++(0,-0.5);

      \filldraw (0,0) circle (3pt) node (l){};
      \filldraw (2,0) circle (3pt) node (rm){};

      \draw [<-] (a) --  (b);
      
      \draw [<-] (b) -| (l);
      \draw [<-] (a) -| (l);
      \draw [<-] (l) -- (-0.4,0);

      \draw [<-] (rm) |- (b);
      \draw [<-] (rm) |- (a);
      \draw [<-] (2.4,0) -- (rm);
    \end{tikzpicture}}
  \caption{Product of two MPOs for the tensors on a single site
    $i$. The product is built the same way on all sites $i \in [1, L]$
    of the system. Matching physical indices are contracted and the
    two left and right MPO bond indices merged into one on each side.}
  \label{fig:arithmetic:prod}
\end{figure}

Given two operators $\hat A$, $\hat B$ and their MPO representation
tensors $\{A_i\}$ and $\{B_i\}$, the product $\hat R = \hat A \hat B$
(read from right-to-left, $\hat B$ is applied first) can be built on
each site individually. It is graphically represented in
Fig.~\ref{fig:arithmetic:prod}. The lower physical index of each $A_i$
is contracted with the upper physical index of the corresponding
$B_i$. The left and right MPO indices of the tensors are merged into
one fat index. This procedure results in a MPO with bond dimensions
$w_i^r = w_i^a \cdot w_i^b$. Specifically, the product of two
single-site operators (MPO bond dimension 1) is again a MPO with bond
dimension 1. The scalar products of operators occuring during the
implementation of non-abelian symmetries in tensor networks can
similarly be implemented independently of the operator at hand.

\subsection{\label{sec:arithmetic:add}Sums of Matrix Product Operators}
The sum of two operators $\hat A + \hat B = \hat R$, represented by
MPO components $\{A_i\}, \{B_i\}$ and $\{R_i\}$ can also be
constructed. Considering only the MPO bond indices, i.e.\ treating
$\{A_i\}$ as \emph{matrices} of operators, the components of the
resulting MPO are built as follows:
\begin{align}
  R_1 & = \begin{pmatrix} A_1 & B_1 \end{pmatrix}  \\
  R_{1 < i < L} & = \begin{pmatrix} A_i & 0 \\ 0 & B_i \end{pmatrix} \\
  R_L & = \begin{pmatrix} A_L \\ B_L \end{pmatrix}
\end{align}
For the example of a $L=3$ MPO, it is easy to verify that this
results in the desired form representing $\hat A + \hat B$. The sum of
two MPOs has a bond dimension $w_i^r = w_i^a + w_i^b$.

\section{\label{sec:trunc}Matrix Product Operator Compression}

When constructing the MPO representation of a single-site operator as
described in section \ref{sec:sso}, the resulting operator will have
bond dimension 1 and will be in its most efficient
representation. Products of such single-site operators (such as $\hat
c^\dagger_i \hat c^\dagger_k \hat c_l \hat c_j$) will keep the bond
dimension at 1. However, the bond dimension will grow linearly in the
number of such terms that are added together. Naively, a four-term
interaction MPO representing $\sum^L_{ijkl} \hat c^\dagger_i \hat
c^\dagger_k \hat c_l \hat c_j$ will have a maximal bond dimensions
$w_{L/2} = O(L^4)$. The leading term in the computational cost of DMRG
typically scales linearly in the maximal $w_i$ and linearly in $L$,
but there are sub-leading terms of quadratic order in $w_i$. Hence,
some way to avoid this quintic or even decic scaling is absolutely
necessary.

\emph{Compressing} a MPO will in general reduce its bond dimension to
the bare minimum. For example, the sum of two identical MPOs will have
a doubled bond dimension which is obviously not necessary -- a
prefactor of $2$ multiplied into the first tensor would correspond to
the same operator. Similarly, two addends with long strings of
identities to the left and right of the active sites, such as $\hat
n_i + \hat n_{i+1}$ can easily ``share'' these strings such that the
most efficient MPO has bond dimension 1 everywhere but on bond
$(i,i+1)$, where 2 is the minimum required.

The compression methods presented here for MPOs are based on the same
idea as those for MPS: Given a MPO which has components $W_i$ and
$W_{i+1}$ on sites $i$ and $i+1$, it is possible to rewrite
\begin{align}
  W_i \to W^\prime_{i} & := W_i p \\
  W_{i; w_{i-1} w^\prime_i}^{\prime \sigma_i \tau_i} & := \sum_{w_i} W_{i; w_{i-1} w_i}^{\sigma_i \tau_i} p_{w_i w^\prime_i} \\
  W_{i+1} \to W^\prime_{i+1} & := p^{-1} W_{i+1} \\
  W_{i+1; w^\prime_{i} w_{i+1}}^{\prime \sigma_{i+1} \tau_{i+1}} & := \sum_{w_i} p^{-1}_{w^\prime w_i} W_{i+1; w_{i} w_{i+1}}^{\sigma_{i+1} \tau_{i+1}} 
\end{align}
without changing the MPO itself. For some MPO components, it is
possible to find matrices $p \in \mathbb{C}^{w_i \times w_i^\prime}$
with $w_i^\prime < w_i$. The new tensors $W^\prime$ then have a
smaller bond dimension $w^\prime_i$ while representing the same
original MPO, as only the matrix product of $W_i$ and $W_{i+1}$ or
$W_i^\prime$ and $W_{i+1}^\prime$ is relevant for the operator. This
is entirely analogous to MPS, which also offer this gauge freedom and
where it is also possible to use it in order to compress the size of
the MPS.

It must be stressed that the compression methods presented here work iteratively
on a bond-by-bond basis and cannot find a globally different (but
better) MPO representation. However, for MPOs investigated here, we
are still able to recover the optimal representation in most cases and
a near-optimal representation even for extremely difficult
problems. For the latter, it would be possible to combine the
compression methods here with others, such as an iterative fitting
method.\cite{froewis10:_tensor}

\subsection{\label{sec:trunc:svd}Rescaling Singular Value Decomposition}
The singular value decomposition of MPOs has been proposed
before\cite{froewis10:_tensor, schollwoeck11} and in
infinite-precision arithmetic it would work exactly the same as for
MPS: Given a tensor ${W}_{i; w_{i-1} w_{i}}^{\sigma_i \tau_i}$, the
indices $w_{i-1}, \sigma_i$ and $\tau_i$ are combined into a larger
index $\gamma$, yielding the matrix $M_{\gamma w_i}$. This matrix is
decomposed via SVD as
$M_{\gamma w_i} = U_{\gamma w^\prime_i} \cdot S_{w^\prime_i
  w^\prime_i} \cdot V_{w^\prime_i w_i}$.
Columns of $U_{\gamma w^\prime_i}$ and rows of $ V_{w^\prime_i w_i}$
which correspond to negligible singular values in
$S_{w^\prime_i w^\prime_i}$ are removed. $U_{\gamma w^\prime_i}$ is
reshaped into the compressed tensor
${W^\prime}_{i; w_{i-1} w^\prime_{i}}^{\sigma_i \tau_i}$.  The product
$S_{w^\prime_i w^\prime_i} \cdot V_{w^\prime_i w_i}$ acts as a
transfer matrix and is multiplied into the next tensor on the right,
compressing the dimension of the MPO on bond $(i, i+1)$. Sweeping
left-to-right and right-to-left through the MPO compresses all bonds.

\begin{figure}
  \centering
  \includegraphics[width=\columnwidth]{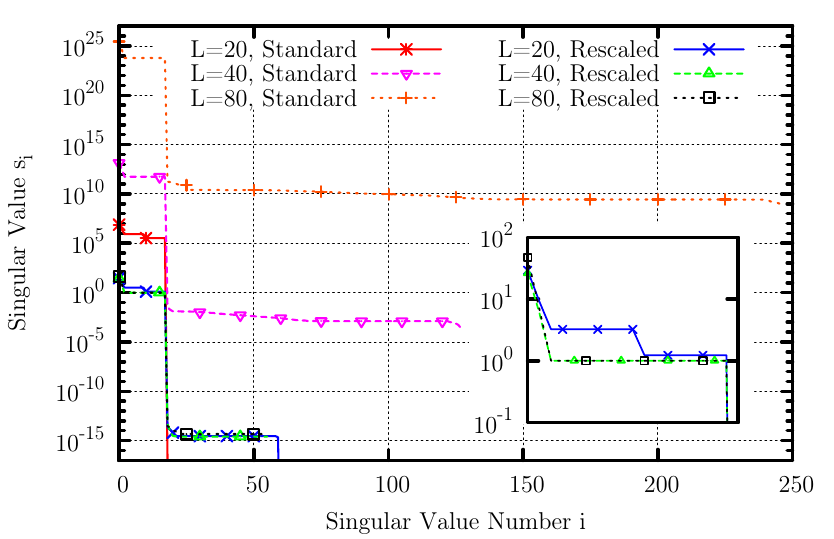}
  \caption{(Color online) Typical singular value distribution observed
    during the compression of a large Hamiltonian MPO with (Rescaled)
    and without (Standard) rescaling for different system sizes. There
    is a sharp drop in magnitude of the singular values between those
    relating to components of the operator and those made redundant by
    the SVD. Without rescaling, the overall magnitude of singular
    values grows exponentially with system size, leading to numerical
    errors. With rescaling, all relevant singular values have
    magnitude independent of the system size. \emph{Inset:} Zoom-in on
    the relevant first 18 singular values with rescaling, which are of
    roughly constant magnitude independent of the system size.}
  \label{fig:examples:svd}
\end{figure}

Unfortunately, a straightforward SVD yields extremely large singular
values. The issue can be observed in Fig.~\ref{fig:examples:svd}
(labels ``Standard'').  Given an uncompressed Fermi-Hubbard Hamiltonian
with some finite range interaction on systems of length $L$ equal to
20, 40 and 80, we compress the left and right halves and then
calculate the singular value spectrum in the centre of the
system. With increasing system size, we observe singular values
growing as large as $10^{26}$. At the same time, the numerical noise,
singular values normally discarded, grows as large as $10^{12}$!

The magnitude of the singular values is linked to the fact that while
we can always rescale a MPS to have Frobenius norm 1, an operator will
in principle have a system-size dependent Frobenius norm. Normalising
all tensors but one, as is common for SVD, implies that only this one
tensor will carry the full norm of the operator.

This leads to two problems: First, there is difficulty in decididing
which singular values should be kept, as even the singular values
strictly associated to numerical noise become extremely large. Second,
compared to the normalised tensors to its left and right, the entries
of the singular value tensor will have a grossly different order of
magnitude, resulting in great precision loss during subsequent
operations.

To avoid such large singular values, we can \emph{rescale} the
singular value tensor $S$ by a scalar value. While this destroys the
ortho\emph{normality} of the resulting MPO bond basis, it preserves
the ortho\emph{gonality}. Further, since all basis vectors are still
of the same length, compression can proceed as usual, either based on
a sharp cut-off or on a dynamic detection of the drop-off in magnitude
of singular values (cf.~Fig.~\ref{fig:examples:svd}). Lastly, properly
chosen, such a rescaling can most often ensure that the norm of the
operator is evenly distributed throughout its length, rather than
concentrated in a single place.

In practice, we found it helpful to calculate the arithmetic average
$a_S$ of the singular values in the tensor $S$ and rescale $S \to
\frac{1}{a_S} S$ such that this average is of order one. The tensor
$W_i^\prime$ is multiplied with the inverse of the scaling factor to
preserve the overall norm. To minimize numerical instabilities, it is
advisable to choose the power of two closest to $a_S$ as the scaling
factor, since such multiplications are exact with IEEE-754 floating
point numbers.

With this rescaling after each SVD during the compression sweeps, we
observe singular values of magnitudes between 1 and 100 independent of
the system length and numerical noise clearly recognisable as such of
magnitude $10^{-14}$ or smaller.

However, there are still caveats and counterindications against using
SVD in specific cases, primarily concerning MPO representations of
projectors or sums of operators involving projectors. First, when
attempting to compress a suboptimal representation of a projector, SVD
even with rescaling often struggles to properly distribute the norm
throughout the system. For example, given the projector
\begin{equation}
  \hat P_\downarrow = \prod_{i=1}^L \left( \frac{\hat 1_i}{2}  - \hat s^z_i \right)
\end{equation}
on the $S=\frac{1}{2}$ Heisenberg chain of lenght $L$, a SVD
compression will lead to exponentially large terms in the first and
last tensor, with the (otherwise properly compressed) terms in the
bulk all carrying a prefactor $\frac{1}{2}$.

Second, when attempting to evaluate sums of operators with greatly
varying Frobenius norms, SVD will often entirely discard the smaller
operator. This is not a concern for most Hamiltonians, as they are
built from few-body interaction terms all with roughly the same order
of magnitude. However, when evaluating $\hat 1 - \hat P_\downarrow$ in
the above system, the result from SVD is simply $\hat 1$. This can be
understood since $||\hat 1||_\mathrm{Frob} = \sqrt{d^L} = 2^{L/2}$
while $||\hat P_\downarrow||_\mathrm{Frob} = 1$. In a similar fashion,
if SVD were tasked with the compression of the sum of two MPS, one of
norm $\sqrt{d^L}$ and the other of norm 1, the result would also
simply be the larger of the two states, as soon as the difference in
the two is lost in the numerical noise of order $10^{-16}$.

Both problems can be detected reliably: For the first, it is
sufficient to compare e.g.~the norm of each MPO component: If one or
two (i.e.~at the edges of the system) is much greater than in the
bulk, SVD failed to properly distribute the norm.

For the second, it is sufficient to compare the Frobenius norms of
addends before operator addition, if in doubt. As a rule of thumb, the
Frobenius norm is exponential in the number of identity MPO
components. It is hence possible to sum few-body interaction terms
together (as they typically occur in Hamiltonians or correlators) or
\emph{alternatively} sum ``few-identity'' terms (such as projectors)
together. However, this rule only applies to sums of MPOs, not
products of MPOs. Special care must be taken in those cases and it
must be checked carefully whether the errors introduced by SVD are
acceptable relative to the problem at hand.

\subsection{\label{sec:trunc:dep}Deparallelisation}
The SVD method has the disadvantage that it destroys the extreme
sparsity of the usual MPO tensors and relies on a robust and small
window of singular values encountered in the MPO. Employing quantum
number labels reduces this destruction of sparsity to the scope of
individual blocks, which will often be implemented as dense tensors in
any case. However, in particular for simple homogenous operators, it
is desirable to keep the sparse, natural structures of MPO. It is
furthermore sometimes also necessary to compress MPOs with greatly
varying singular values.

The much simpler deparallelisation method avoids both issues entirely:
Sparsity is largely conserved and the compression does not rely on
singular values. It furthermore does not rescale most elements of the
tensor, keeping the norm distributed in the same way as before. It was
first presented in Ref.~\onlinecite{mcculloch07:_from} and can be
considered a slight generalisation from the \emph{fork-merge} method
presented in Ref.~\onlinecite{keller15:_hamil}, from ``forking'' and
``merging'' only identity operators to arbitrary strings of operators.

The algorithm is presented in detail in Appendix \ref{app:depara}. The
basic idea is again to re-shape each site tensor
$W^{\sigma_i \tau_i}_{i; w_i w_{i+1}}$ into a matrix
$M_{\gamma w_{i+1}}$. Then, columns of $M$ which are entirely parallel
to any previous column are removed, with the respective
proportionality factor stored in the transfer matrix to be multiplied
into the next site tensor.

This procedure results in a MPO that is often optimal for spatially
homogeneous operators and retains the advantageous structure of
analytically-constructed MPO tensors. For more difficult Hamiltonians,
it often results in suboptimal representations.

\section{\label{sec:trunc:del}Delinearisation}
The delinearisation method aims to combine the advantages of the SVD
and the deparallelisation. It is suitable to compress any MPO,
including the previously-mentioned sums of projectors and Hamiltonians
as well as complicated Hamiltonians. In most cases, it results in an
optimal MPO dimension. For extremely large MPOs, the resulting bond
dimensions tend to be slightly larger than with SVD
compression. However, the original sparsity of the MPO is largely
preserved, even in the dense sub-blocks\footnote{When quantum numbers
  (both from Abelian and Non-Abelian symmetries) are employed in
  tensor networks, it becomes possible to split the original, dense
  tensor into smaller dense tensor blocks which transform uniquely
  under these quantum numbers. The size of these blocks, e.g.~the
  number of states with total spin $S=0$, is the most relevant scaling
  dimension.} labelled by quantum numbers. Wherever possible, it
attempts to ensure that no spurious small terms can occur in the
Hamiltonian.

The algorithm is presented in full detail in Appendix
\ref{app:delin}. Similar to the deparallelisation, we attempt to
remove columns from the $M_{\gamma w_{i+1}}$ matrix, but now allow for
linear combinations of previously-kept columns to replace the column
in question, whenever possible under the constraint that no
cancellation to exactly-zero can occur (this avoids the spurious small
terms).

\section{\label{sec:example}Examples}
We will present three examples of MPO generation using the above
construction method. First, we show that it works well for the simple
example of nearest-neighbour interactions on spin chains and even for
small powers of the Hamiltonian. Second, we explain that it is very
easy to generate the Hamiltonian for the Fermi-Hubbard model on a
cylinder in hybrid real- and momentum space. Third, we present data
that the construction method also correctly sums up partial terms in a
toy model for the full quantum chemistry Hamiltonian.

\subsection{Spin Chains with Nearest-Neighbour-Interactions}
\begin{figure}
  \centering
  \includegraphics[width=\columnwidth]{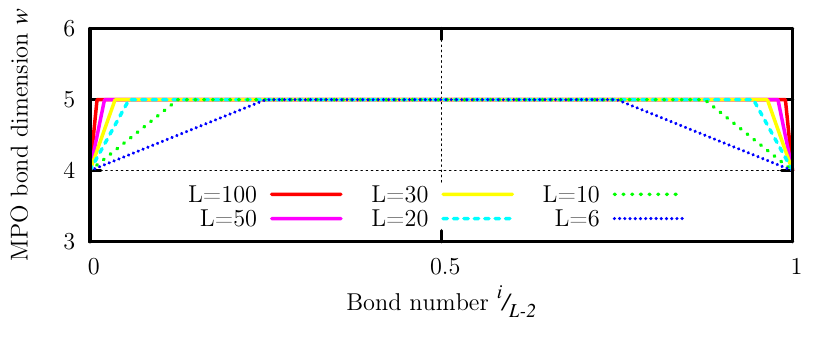}
  \caption{(Color online) MPO bond dimension of the representation of
    Eq.\ \eqref{eq:example:spin:H} over the length of the chain for
    different chain lengths, here at $h=1$, $J_x = \frac{1}{2}$, $J_y
    = \frac{1}{3}$ and $J_z = \frac{1}{5}$ to illustrate the generic
    case. The leftmost and rightmost bonds have dimension four,
    whereas the bulk bond dimension is five as in the analytical
    solution.}
  \label{fig:example:spin:w}
\end{figure}

We consider the Hamiltonian with nearest-neighbour interactions on a
spin chain
\begin{align}
  \hat H = & \sum_{i=1}^L h \hat S^z_i + \sum_{a=x,y,z} \sum_{i=1}^{L-1} J_a S^a_i S^a_{i+1}\label{eq:example:spin:H}.
\end{align}
We can construct analytically an optimal representation of this
Hamiltonian,\cite{schollwoeck11} which has MPO bond dimension 5. In
comparison, we can plot the dimension of each bond of the numerically
constructed representation for various system sizes
(cf.~Fig.~\ref{fig:example:spin:w}). As is clearly visible, the bond
dimension quickly saturates at five and stays constant $O(1)$
independent of the system length. The algorithm even finds an
improvement over the usual analytic solution, as only one $\hat S^z$
term is necessary at the boundary. In the bulk, it completely
reproduces the analytic solution, here at the example of $J_x = J_y =
J_z = h = 1$:
\begin{equation}
  {W}_{\textrm{bulk}} =
  \begin{pmatrix}
    1 & s^z & s^z & s^y & s^x \\
    0 & 1 & 0 & 0 & 0 \\
    0 & s^z & 0 & 0 & 0 \\
    0 & s^y & 0 & 0 & 0 \\
    0 & s^z & 0 & 0 & 0
  \end{pmatrix}
\end{equation}

Further, we can construct powers of the Hamiltonian $\hat H$, here
specifically with coefficients $h=1$, $J_x = \frac{1}{2}$, $J_y =
\frac{1}{3}$, $J_z = \frac{1}{5}$. The procedure is to first generate
$\hat H$ using only deparallelisation, which leads to the
near-analytic solution at bond dimension 5. We then multiply the MPO
with itself to generate $\hat H^2$ and compress the operator using SVD
or Delinearisation. Multiplying with $\hat H$ repeatedly, we construct
up to the seventh power of $\hat H$ and compare the bond dimensions
with those resulting from an iterative fitting procedure (cf.\ Table
\ref{tab:example:spinchain}).

For small powers $n$, the resulting bond dimensions from the three
compression methods coincide. For higher powers, the SVD method
results in somewhat lower bond dimensions. This could be both due to
numerical inaccuracies in either method (e.g.\ erroneously discarding
small but relevant singular values) or the fitting approach getting
stuck in a local minimum. To numerical accuracy, the error resulting
from the SVD compression is zero, however.

In comparison, the Delinearisation method encounters cyclic linear
dependencies it cannot break when attempting to compress the higher
power MPO representations. This results in a larger bond
dimension. However, the original sparsity of the MPO, calculated as
the relative number of exactly-zero entries in the dense sub-blocks of
the MPO, is largely preserved at over 80\% zero entries while no such
entries where found after SVD compression.

\begin{table}
  \caption{\label{tab:example:spinchain}Bond dimensions $w_{L/2}$ in the center of a $L=100$ chain of powers of the nearest-neighbour spinchain Hamiltonian \eqref{eq:example:spin:H} with SVD and Delinearisation compression. Relative sparsity of the resulting MPO is included for the Delinearisation method (SVD does not preserve sparsity at all). We compare with the results of Fröwis et.\ al.\ \cite{froewis10:_tensor} for the XXZ Hamiltonian constructed with an iterative fitting procedure, which could also be combined with our construction method for MPOs.}
  \begin{ruledtabular}
  \begin{tabular}{l|ccccccc}
    Order $\hat H^n$ & 1 & 2 & 3 & 4 & 5 & 6 & 7 \\
    \hline\hline
    SVD: $w_{L/2}$  & 5 & 9 & 16 & 32 & 51 & 64  & 92 \\
    DLN: $w_{L/2}$  & 5 & 9 & 16 & 32 & 51 & 81 & 126 \\
    DLN: Sparsity & 81\% & 84\% & 82\% & 89\% & 88\% & 88\% & 85\% \\
    Fitting Method:\cite{froewis10:_tensor} $w_{\mathrm{max}}$ & 5 & 9 & 16 & 32 & 51 & 79 & 110 \\
  \end{tabular}
\end{ruledtabular}
\end{table}

\subsection{Fermi-Hubbard Hamiltonian on a Cylinder in Hybrid Space}
The Fermi-Hubbard model in two dimensions is a problem of ongoing
research. When attempting a solution of two-dimensional problems with
DMRG, the usual course of action is the mapping onto a
cylinder.\cite{stoudenmire12:_study_two_dimen_system_densit} This
allows for periodic boundary conditions along the cylinder width,
while keeping the ends of the cylinder open, as is advantageous for
DMRG. With coordinates $x$ along the length $L$ of the cylinder and
coordinates $y$ along its width $W$, we can then write the Hamiltonian
as
\begin{align}
  \hat H = - & \sum_{y=1}^W \sum_{x=1}^L \hat c^\dagger_{x,y} \cdot \hat c_{x,y+1} + \mathrm{h.c.} \nonumber \\
           - & \sum_{y=1}^W \sum_{x=1}^{L-1} \hat c^\dagger_{x,y} \cdot \hat c_{x+1,y} + \mathrm{h.c.} \nonumber \\
           + & \frac{U}{2} \sum_{y=1}^W \sum_{x=1}^L \left(\hat c^\dagger_{x,y} \cdot \hat c_{x,y}\right)^2 - \hat c^\dagger_{x,y} \cdot \hat c_{x,y} \quad.
\end{align}                       

The first two lines are the usual kinetic term containing
nearest-neighbour hopping along the width and the length of the
cylinder. The third term is the on-site interaction with coefficient
$U$, already written in anticipation of the following Fourier
transformation. The objects $c^\dagger$ and $c$ are
$\mathrm{SU}(2)$-invariant operators, the scalar product $c^\dagger
\cdot c$ could be expanded out to the usual form $c^\dagger_\uparrow
c_\uparrow + c^\dagger_\downarrow c_\downarrow$. This model has
explicit $\mathrm{SU}(2)_{\mathrm{Spin}}$ symmetry with quantum number
$S$ and $\mathrm{U}(1)_{\mathrm{Charge}}$ symmetry with
quantum number $N$.

Following Ref.~\onlinecite{motruk16:_densit}, we can perform a Fourier
transformation along the width of the cylinder with the identities
\begin{align}
  \hat c_{x,y} & = \frac{1}{\sqrt{W}} \sum_{k=1}^W e^{2 \pi \mathrm{i} \frac{k}{W} y} \hat c_{x,k} \\
  \hat c^\dagger_{x,y} & = \frac{1}{\sqrt{W}} \sum_{k=1}^W e^{-2 \pi \mathrm{i} \frac{k}{W} y} \hat c_{x,k}
\end{align}
to achieve the form
\begin{align}
  \hat H = - \sum_{x=1}^L \sum_{k=1}^W &  2 \cos\left(2\pi \frac{k}{W}\right) \hat c^\dagger_{x,k} \cdot \hat c_{x,k} \nonumber \\
           - \sum_{x=1}^{L-1} \sum_{k=1}^W &  \hat c^\dagger_{x,k} \cdot \hat c_{x+1,k} + \mathrm{h.c.} \nonumber \\
           + \frac{U}{2} \sum_{x=1}^L \sum_{k=1}^W &  \left[ \sum_{lm=1}^W \frac{1}{W} \left(\hat c^\dagger_{x,k} \cdot \hat c_{x,l} \right) \left( \hat c^\dagger_{x,m} \cdot \hat c_{x,k-l+m} \right)\right] \nonumber \\
           &  - \hat c^\dagger_{x,k} \cdot \hat c_{x,k} \quad.
\end{align}

The individual terms are again the width-wise hopping, now diagonal,
then the length-wise hopping which did not change and the interaction
term. The interaction term is now a proper four-body interaction
spanning every ring of the cylinder. For very wide cylinders, this
would be prohibitively expensive, but since DMRG is exponentially
bound in the width of the cylinders due to the entanglement structure
regardless of the choice of real- or momentum space basis, this is not a
major concern.

Again following Ref.~\onlinecite{motruk16:_densit}, we can also
exploit the $\mathbb{Z}_W$ momentum conservation symmetry to increase
the sparsity of both the MPO and the MPS by attaching an additional
quantum number $K$ to each local basis state. This requires a
non-homogenous basis, as the 1-electron state on site $(x,k)$ has to
transform as $K=k$ (while of course keeping its spin and charge
quantum numbers $S$ and $N$, respectively). Similarly, the
two-electron state on this site has to transform as $K=2k$. For the
same reason, the operator $\hat c^\dagger_{x,k}$ and its MPO
representation not only transform as $S=\frac{1}{2}$ and $N=1$ but
also as $K=k$, which depends on the site on which the operator acts.

There is additional freedom in the choice of ordering of the
two-dimensional pairs $(x,k)$ on the one-dimensional DMRG chain. Here,
we choose a Z-like pattern, connecting the last site of each ring to
the first site of the next ring. Further, we have the freedom to
re-arrange the momentum sites $k$ within the ring, ideally to minimise
both the bond dimension of $\hat H$ as well as to reduce the
entanglement of the resulting MPS groundstate. Expecting
antiferromagnetic correlations, we place momenta separated by $\pi$
next to each other, e.g. for 8 sites, the ordering within a ring is $k
= 0, \pi, \frac{\pi}{4}, \frac{5\pi}{4}, \frac{\pi}{2},
\frac{3\pi}{2}, \frac{3\pi}{4}, \frac{7\pi}{4}$.

\begin{figure}
  \centering
  \includegraphics[width=\columnwidth]{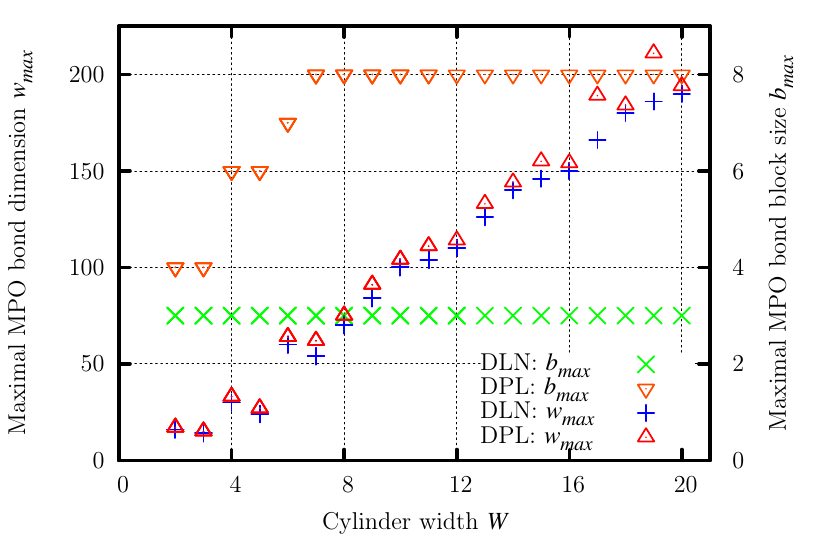}
  \caption{(Color online) Maximal MPO bond dimension (left axis) and
    maximal block size (right axis) for the Fermi-Hubbard Hamiltonian
    on a cylinder with nearest-neighbour hopping and (real-space)
    on-site interactions. The maximal bond dimension occurs in the
    middle of each ring of the cylinder and is independent of the
    cylinder length. This is also the bond on which the size of the
    largest quantum number block becomes maximal if only
    Deparallelisation (DPL) compression is employed. On this problem,
    SVD and Delinearisation (DLN) result in the same bond
    dimensions. With either of the two, the largest quantum number
    block is fairly uniformly three, only dropping down to two on the
    inter-ring connections. }
  \label{fig:example:fh}
\end{figure}

The most important scaling is given by the the maximal bond dimension
of the MPO. The size of this parameter affects the runtime needed by
DMRG. Fig.~\ref{fig:example:fh} shows the maximal bond dimension of
the MPO representation constructed using this method for $U=2$. This
maximal bond dimension is independent of the cylinder length $L$ and
occurs in the middle of each ring. In comparison with the simple
Deparallelisation, the total bond dimensions resulting from either the
SVD compression or the Delinearisation compression are only slightly
smaller. However, both the SVD and Delinearisation method reduce the
size of the largest dense blocks in the tensors from 8 to 3.

Inspecting those dense blocks on e.g.~a $16 \times 4$ lattice, we find
that with the delinearisation method, 8\% of the stored values are
exactly zero, 14\% exactly $-1$ and 32\% exactly $+1$. In comparison,
after SVD compression, an exactly-zero value never occurs and the two
most common values are $\approx \pm \sqrt{4/3}$ at 4\% and 14\%
respectively. Hence even in such small blocks of size at most $3
\times 3$, the Delinearisation method preserves sparsity and a
relatively simple tensor structure to a noticeable degree.

Independent of the compression method, we observe largely linear
growth of the maximal bond dimension with the cylinder width. This can
be explained by the momentum conservation in the interaction term:
given two fixed operators on one half of the system and a third
operator on the other half, there is only one valid location for the
fourth operator. Hence we get overall $O(L)$ scaling.

\subsection{Full Electronic Randomised Fermi-Hubbard}

Contrary to the fairly homogenous problems in solid-state physics, the
application of MPO-based algorithms in quantum chemistry is more
difficult.\cite{keller15:_hamil} In particular, there are often
long-range four-body interactions with different coupling
coefficients. As a toy model for such a Hamiltonian, we consider the
operator
\begin{equation}
  \hat H = \sum_{\sigma \tau=\uparrow\downarrow} \sum^L_{ijkl} V_{ijkl} \hat c^\dagger_{i\sigma} \hat c^\dagger_{k\tau} \hat c_{l\tau} \hat c_{j\sigma} \label{eq:examples:fqc:H}
\end{equation}
with $V_{ijkl} = V_{jilk}$, $|V_{ijkl}| < 2$ but coefficients
otherwise random. Construction of this Hamiltonian with the presented
method is extremely expensive -- $L^4$ MPO-MPO additions have to be
evaluated and the intermediate sums have to be continously compressed
to avoid quartic growth of bond dimensions. Nevertheless, we are able
to construct the MPO representation of the Hamiltonian for $L$ up to
$\approx 34$. With more advanced techniques and a modest amount of
preprocessing, which are outside the scope of this paper, it would be
possible to also construct the Hamiltonian for larger systems.

For this $\hat H$, the maximal bond size always occurs in the middle
of the system at bond $L/2$ (or bonds $L/2 - 1$ and $L/2$ for odd
$L$). It is possible to sum up partial terms in
\eqref{eq:examples:fqc:H} to the left and right of a given bond
s.t. there are only $O(L^2)$ terms remaining on either
side.\cite{keller15:_hamil}

\begin{figure}
  \centering
  \includegraphics[width=\columnwidth]{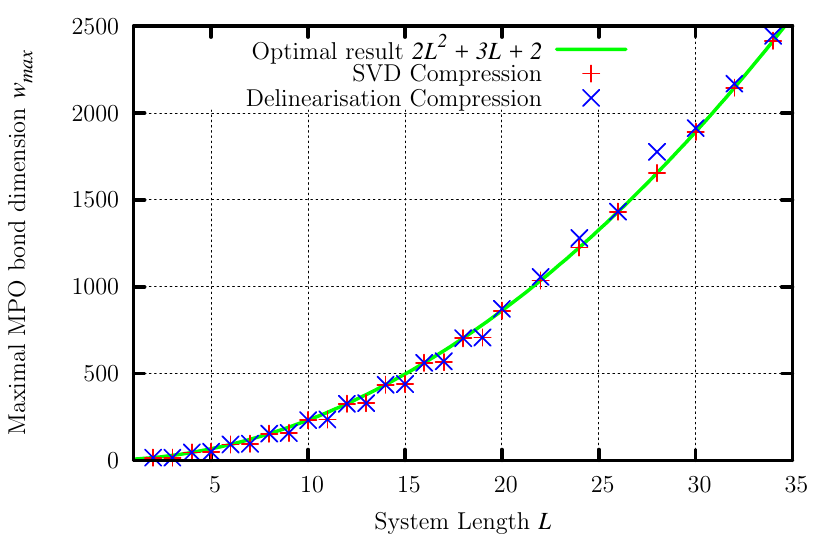}
  \caption{(Color online) Maximal MPO bond dimension in the middle of
    the chain for the representation of the full quantum chemistry
    Hamiltonian in \eqref{eq:examples:fqc:H}. The maximal bond
    dimension after SVD compression for even lengths behaves exactly
    as $w_{\mathrm{max}}(L) = 2 L^2 + 3 L + 2$, which is the optimal
    result. We included some data for odd lengths $L$ for
    completeness; the increase in bond dimension from $L=2n$ to
    $L=2n+1$ is consistently four with SVD. With delinearisation
    compression, the result is exactly the same as with SVD for $L
    \leq 16$, for larger system sizes, the optimal representation is
    not always found, but the scaling is still decidedly quadratic.}
\label{fig:example:fqc:w}
\end{figure}

Fig.~\ref{fig:example:fqc:w} shows the maximal bond dimension for a
given system length after construction from single-site operators via
multiplication, addition and compression by deparallelisation (every
$L$ steps) and SVD (every $L^2$ steps).  Using the Delinearisation
method instead of the SVD compression, the resulting bond dimensions
increase slightly at larger system sizes, as the Delinearisation
cannot always break cyclic linear dependencies in its input
columns. However, the bond dimensions as returned by Delinearisation
still have decidedly quadratic scaling.

The main advantage of the method is then in its flexibility: Adding
another type of interaction, changing coefficients or changing the
system size can be done independently of the implementation of the
compression methods as well as the definition of the single site
operators.

The parameters of the optimal result can be explained as follows:
First, there are always two identity terms which correspond to
summands with all $i, j, k, l$ to the left or right of the center
bond, resulting in the two constant terms. Further, there are $L$
contributions each for one out of $i, j, k, l$ to the left and three
to the right (and vice versa) as well as $L$ contributions for $i = j$
or $k = l$. Finally, there are $L^2$ ways each to distribute two out
of $i, j, k, l$ on the left or the right of the system. This is in
agreement with recent results by Chan et.\ al.\
\cite{kin-lic16:_matrix_produc_operat_matrix_produc} who also find a
leading term $2 L^2$. The SVD compression therefore leads to the optimal
MPO representation with scaling $O(L^2)$.

\section{\label{sec:variance}Calculation of higher moments}

\begin{figure}
  \centering
  \includegraphics[width=\columnwidth]{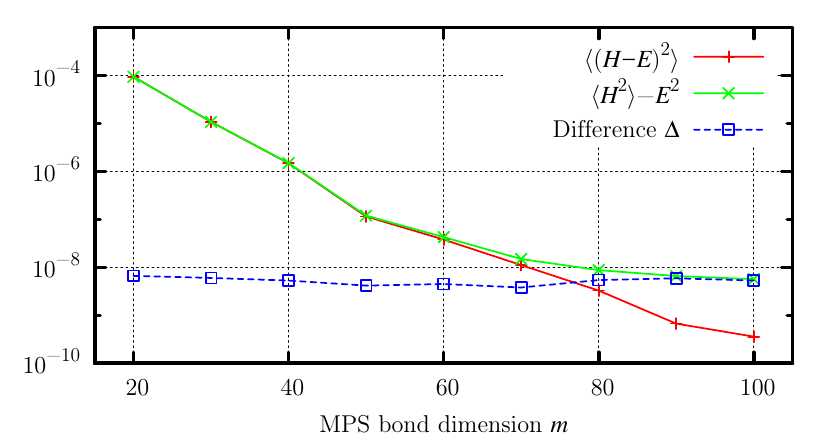}
  \caption{(Color online) The variance $\langle (H - E)^2 \rangle$ of
    the MPS approximation to the groundstate of the $S=1$ Heisenberg
    chain, as a function of bond dimension $m$. The naive calculation
    of $\langle H^2 \rangle - E^2$ is subject to catastrophic
    cancellation and cannot be obtained accurately.  The properly
    constructed MPO representation for $(H-E)^2$ is well-conditioned
    and obtains full numerical precision. The difference $\Delta =
    \left(\langle H^2 \rangle - E^2\right) - \left\langle (H - E)^2
    \right\rangle$ is consistently of order $10^{-8}$, i.e.~relevant
    as soon as the variance becomes sufficiently accurate.}
  \label{fig:Variance}
\end{figure}

The calculation of higher moments is an obvious application of MPO
techniques.  For example, the energy variance $\sigma^2 = \langle H^2 \rangle - \langle H \rangle^2 = \langle (H-E)^2 \rangle$
is a robust alternative to calculating the truncation error that has
many advantages.\cite{mcculloch07:_from} However, the naive
computation of the variance as the difference between the second
moment $\langle H^2 \rangle$ and the square of the energy is extremely
prone to catastrophic cancellation\cite{Higham:2002:ASN} due to
subtraction of two numbers that have a large magnitude, whereas the
result has typically a small magnitude. In double-precision floating
point numerics that are typical for MPS calculations, there are
approximately 16 decimal digits of precision available, so if one
wants to be able to resolve a variance of, say $10^{-10}$, this
implies that the total energy of the system can be no larger than
$10^3$.  In practice that is a gross overestimate, since roundoff
errors will account for at least a digit or two as well, which means
that in typical calculations one encounters numerical problems
evaluating the variance when the system size gets to around
$L \sim 100$ or so sites.  The solution to recovering numerical
precision is to construct an MPO representation of $(H-E)^2$ directly,
thereby distributing the constant energy term across each site of the
MPO. The intermediate sums formed when contracting the MPO are then
bounded to be $O(1)$ -- the only component of the summation that
diverges with system size is the variance itself, which is only linear
in $L$.  The MPO representation $(H-E)^2$ is straightforward to
construct, by starting from the MPO representation of $H$ and
subtracting the local contribution to the energy at each site,
\begin{equation}
W^{H-E}_i = W^{H}_i - E_i \; ,
\end{equation}
where each $E_i$ is the contribution to the energy due to site
$i$. Unless a better value is available, $E_{\mathrm{total}}/L$ can be
used here. The squared MPO can then be generated straightforwardly,
and it is important that the MPO compression scheme preserves the
structure of the MPO, which is true at least for the parallel
compression algorithm.  This loss of accuracy through catastrophic
cancellation is demonstrated in Fig.~\ref{fig:Variance}.  This is
obtained for a spin $S=1$ Heisenberg chain with $L=100$ sites.  With
the uncontrolled algorithm, catastrophic cancellataion limits the
accuracy of the variance calculation to $O(10^{-8})$, but with proper
construction of the MPO, the variance can be calculated to full
accuracy. This difference in how the variance is computed depends
drastically on the system size, since the variance is linear in system
size but the cancellation of terms in $\langle H^2 \rangle - E^2$ is
$O(L^2$).  Higher moments are affected even more drastically, since
the $k^\mathrm{th}$ moment involves subtraction of terms of order
$L^k$. The generalization to higher moments is best viewed in terms of
the cumulant expansion, since each cumulant is linearly extensive in
the system size and they exactly capture the numerical divergences of
the moment expansion. The first cumulant $\kappa_1$ is just the energy
itself, and the second cumulant $\kappa_2 = \sigma^2$ is the variance.
The $3^\mathrm{rd}$ cumulant, which characterizes the skewness of the
distribution, is given by
$\kappa_3 = \langle H^3 \rangle - 3 \langle H^2 \rangle \langle H
\rangle + 2 \langle H \rangle^3$,
and is obtained from the MPO representation
\begin{equation}
W^{[3]} = \left(\left(H - \kappa_1 \right)^2 - \kappa_2 \right)\left(H-\kappa_1\right) \; .
\end{equation}
If the MPO is properly constructed in this way then there are no intermediate terms
that grow with the system size as the MPO is contracted, 
and there is essentially no practical limit to the accuracy
of evaluating higher order moments (the exponential time cost from the dimension of the MPO
is the limiting factor, not accuracy of computation).

This structure is implicit in the triangular MPO formulation for
infinite systems.\cite{idmrg}  In determining the expectation values
of higher moments of an iMPO, the recursive formulation presented in
Ref.~\onlinecite{idmrgMoments} keeps control over the numerical
precision without explicitly removing the energy contributions, due to
the particular triangular structure (Jordan form) of the expectation
values. The contributions to the moment that diverge with each power
of the system size are obtained separately as the coefficients of a
polynomial expansion.  Efficient compression of translationally
invariant infinite MPOs has some distinct features compared with
finite MPOs, and this will be described in detail in a future
publication.\cite{impoOptimization}

\section{\label{sec:conclusions}Conclusions \& Outlook}
This paper presents a generic construction method to generate an
efficient MPO representation of arbitrary operators in the context of
second-generation DMRG algorithms. The method only requires the
definition of single-site operator tensors and the implementation of
arithmetic operations on MPOs as well as compression of a MPO. Any
operator can then be expressed in a few loops of any object-oriented
programming language. In turn, the method facilitates the study of
varying and complex systems, as the amount of work to be done up front
prior to DMRG calculations is substantially lowered.

The simplest compression method presented (Deparallelisation) can
handle most nearest-neighbour Hamiltonians, while for more complicated
MPOs, either the SVD or the Delinearisation should be used (cf.\ Table
\ref{tab:comparison-all}).

The resulting MPO either exactly reproduces the optimal analytical
solution (for spatially homogenous short-range operators) or is the
optimal representation which would be difficult to construct
analytically (for medium-range Hamiltonians in two dimensions as well
as powers of short-range Hamiltonians). In principle, it is also
possible to apply the method to the quantum chemistry Hamiltonian, but
computational costs of a naive implementation become too large to be
feasible. It would be possible to combine the compression techniques
presented here with other compression methods in the future, if
necessary. The implementation is compatible with non-abelian spin and
charge symmetries enforced on the tensor level.  The generalisation to
tree-tensor networks is also straightforward.

\begin{table*}
  \caption{\label{tab:comparison-all}Overview comparison of the three compression methods presented in this paper. Computational cost is a rough statement regarding the relative costs of the methods, as all three scale cubically in the bond dimension of the MPO}
  \begin{ruledtabular}
  \begin{tabular}{l|ccccc}
    Method & Optimal $w_{\mathrm{max}}$ & Sparsity & Spurious terms & Implementation Complexity & Computational Cost \\
    \hline \hline
    Deparallelisation & Easy MPOs & Preserved & None & Simple & Cheap \\
    Rescaled SVD & Always & Lost & Yes & Simple (with LAPACK) & Expensive \\
    Delinearisation & Most MPOs & Preserved & Nearly none & Medium & Medium
  \end{tabular}
\end{ruledtabular}
\end{table*}
\begin{acknowledgments}
  C.~Hubig acknowledges funding through the ExQM graduate school and
  the Nanosystems Initiative Munich. I.~McCulloch acknowledges support from
  the Australian Research Council (ARC) Centre of Excellence for Engineered 
  Quantum Systems, grant CE110001013. and the ARC Future Fellowships scheme, FT140100625.
\end{acknowledgments}

\appendix

\section{\label{app:compression}Suggested MPO Compression Procedure}

MPO compression of an arbitrary operator should occur in three stages:

\begin{enumerate}
\item Performing one full sweep using the deparallelisation method
\item Performing sweeps using the strict delinearisation method until bond dimensions stay constant
\item Performing sweeps using the relaxed delinearisation method until bond dimensions stay constant
\end{enumerate}

The motivation for this sequence is to firstly reduce the bond
dimension as much as possible with the fairly cheap deparallelisation,
then move on to the more costly delinearisation and finally, if a
cyclic dependency occurs which cannot be broken without allowing
cancellation to zero, use the relaxed delinearisation. Note that if
the MPO is already optimal, the last step will not introduce such
small terms.

Independent of the compression method, each full sweep iterates twice
over the full system, once from left to right and then from right to
left. On each site $i$, the local tensor $W^{\sigma_i \tau_i}_{i; w_{i-1}
  w_i}$ is re-shaped into a matrix $M_{\gamma w_i}$ $\left(
  M_{\gamma w_{i-1}} \right)$ during left-to-right (right-to-left)
sweeps. The matrix $M$ is then decomposed as $M = \tilde M \cdot T$. $\tilde
M$ is re-shaped into the new site tensor ${\tilde W}^{\sigma_i
  \tau_i}_{i; w_{i-1} \tilde w_i}$ $\left({\tilde W}^{\sigma_i
    \tau_i}_{i; \tilde w_{i-1} w_i}\right)$ with the transfer matrix
$T$ being multiplied into the next site tensor $W_{i+1}$ ($W_{i-1}$)
during left-to-right (right-to-left) sweeps.

The decomposition $M \to \tilde M \cdot T$ is described in the
following sections for the deparallelisation and delinearisation
methods.

\section{\label{app:depara}Deparallelisation Algorithm}

\emph{Input}: Matrix $M_{a b}$

\emph{Output}: Matrices $\tilde M_{a \beta}$, $T_{\beta b}$ s.t.\
$M_{a b} = \sum_{\beta} M_{a \beta} T_{\beta b}$ and $\tilde M$ has at
most as many columns as $M$ and no two columns which are parallel to
each other.

\emph{Procedure}:
\vspace{-0.25cm}
\begin{enumerate}[label=(\arabic*)]
\item Let $K$ be the set of kept columns, empty initially
\item Let $T$ be the dynamically-resized transfer matrix
\item For every column index $j \in [1, b]$:
  \begin{enumerate}[label=(\arabic{enumi}.\arabic*)]
  \item For every kept index $i \in [1, |K|]$:
    \begin{enumerate}[label=(\arabic{enumi}.\arabic*)]
    \item If the $j$-th column $M_{:j}$ is parallel to column $K_i$:
      \begin{enumerate}[label=(\arabic{enumi}.\arabic*)]
      \item set $T_{i,j}$ to the prefactor between the two columns
      \end{enumerate}
    \item Otherwise:
      \begin{enumerate}[label=(\arabic{enumi}.\arabic*)]
      \item add $M_{:j}$ to $K$, set $T_{|K|, j} = 1$.
      \end{enumerate}
    \end{enumerate}
  \end{enumerate}
\item Construct $\tilde M$ by horizontally concatenating the columns
  stored in $K$.
\item Return $\tilde M$ and $T$
\end{enumerate}

The check for parallelicity is ideally done on an element-wise basis
by finding the first non-zero element of either column, calculating
the factor between it and the corresponding element of the other
column and then ensuring that all other elements agree on that
prefactor. Zero columns should be removed with a corresponding zero
column stored in $T$.

\section{\label{app:delin}Delinearisation Algorithm}
\emph{Input}: Matrix $M_{a b}$, threshold matrix $\Delta_{a b}$

\emph{Output}: Matrices $\tilde M_{a \beta}$, $T_{\beta b}$ s.t.\
$M_{a b} = \sum_{\beta} M_{a \beta} T_{\beta b}$ and $\tilde M$ has at
most as many columns as $M$ and all columns in $\tilde M$ are linearly
independent.

\emph{Remark}: Initially, the threshold matrix $\Delta_{a b}$ is
constructed from $W^{\sigma_i \tau_i}_{i; w_{i-1} w_i}$ as
$\Delta_{(\sigma_i \tau_i w_{i-1}) w_i} = \sum_{\sigma_i^\prime
  \tau_i^\prime} |W^{\sigma_i^\prime \tau_i^\prime}_{i; w_{i-1} w_i}|
\cdot \varepsilon_D$,
i.e.\ each element is the 1-norm of the original operator to which it
belongs multiplied by a small threshold.

\emph{Procedure}:
\vspace{-0.25cm}
\begin{enumerate}[label=(\arabic*)]
\item \label{alg:dln:relaxed} If relaxed delinearisation: Set all elements $\Delta_{ab} \equiv 0$ to $\varepsilon_D$.
\item \label{alg:dln:inner:start} Deparallelise the rows of $M_{ab}$:
  \begin{eqnarray}
    M_{ab} \to & R_{a \alpha} M^p_{\alpha b} \\
    \Delta_{ab} \to & R_{a \alpha} \Delta^p_{\alpha b}
  \end{eqnarray}
  where the elements of $\Delta^p$ are chosen as the smallest
  elements in that column from non-zero rows which were parallel to
  the kept row.
\item \label{alg:dln:permutation} Sort the columns of $M^p$ according
  to the following criteria, resulting in $M^{pP}$, $\Delta^{pP}$ and
  a permutation matrix $P$. Sorting criteria are:
  \begin{enumerate}[label=(\arabic{enumi}.\arabic*)]
  \item The number of exactly-zero values in the column
  \item if tied, the number of exactly-zero thresholds in the same
    column of $\Delta^p$
  \item if tied, the number of exactly-zero values from the bottom of
    the column
  \item if tied, the number of exactly-zero thresholds from the bottom
    of the same column of $\Delta^p$
  \end{enumerate}
\item For every column $\mu$ and associated threshold column $\delta$ in $M^{pP}$ and $\Delta^{pP}$
  \begin{enumerate}[label=(\arabic{enumi}.\arabic*)]
  \item Attempt to solve
    \begin{equation} A x = \mu \end{equation} where $A$ is the matrix
    from eligible previously-kept columns. A column is eligible for
    inclusion in $A$ if it has no non-zero entry in a row where
    $\delta$ is exactly zero.

    The coefficients $x$ are found via QR decomposition with column
    scaling (by their respective norms). Rows of $R$ and $Q^H \mu$ are
    scaled s.t. the right-hand side is either 1 or 0 prior to
    solution by backwards substitution.

  \item If any coefficients in $x$ have absolute value less than
    $\varepsilon_t$, remove the associated column from the eligible
    set to build $A$ and repeat.
  \item If any coefficients in $x$ are close to $\pm 1$, replace them
    by $\pm 1$.
  \item If each element $(Ax - c)_i$ of the residual is
    smaller than $\delta_i \times \mathrm{cols}(A)$:
    \begin{enumerate}[label=(\arabic{enumi}.\arabic*)]
    \item store the coefficients $x$
    \end{enumerate}
  \item Else,
    \begin{enumerate}[label=(\arabic{enumi}.\arabic*)]
    \item add the column to the set of kept columns and
      store a coefficient of 1 in the appropriate place.
    \end{enumerate}
  \end{enumerate}
\item Collect all kept columns into $M^{pC}$, associated columns from
  $\Delta^{pP}$ into $\Delta^{pC}$ and construct the transfer matrix
  $T^C$ from the stored coefficients times the permutation matrix $P$.
\item \label{alg:dln:inner:end} Multiply the row-deparallelisation
  transfer matrix $R$ back into $M^{pC}$ and $\Delta^{pC}$, yielding
  $M^C$ and $\Delta^C$.
\item \label{alg:dln:reset} If the number of columns in $M^C$ is equal to the number of
  columns in $M$, replace $M^C = M$, $\Delta^C = \Delta$, $T^C =
  \mathbf{1}$.
\item \label{alg:dln:rows} Repeat steps \ref{alg:dln:inner:start}
  through \ref{alg:dln:inner:end} for $M^{C\dagger}$ and
  $\Delta^{C\dagger}$ (i.e. delinearise the \emph{rows} of $M^C$):
  \begin{eqnarray}
    M^{C\dagger} & = M^{CR} T^{R} \\
    M^C & = T^{R\dagger} M^{CR\dagger}
  \end{eqnarray}
\item \label{alg:dln:reset2} If neither $T^{R\dagger}$ nor $M^C$ have
  fewer columns than $M$
  \begin{enumerate}[label=(\arabic{enumi}.\arabic*)]
  \item return $\tilde M = M$ and $T = \mathbf{1}$.
  \end{enumerate}
\item \label{alg:dln:rows2} Else-If $T^{R\dagger}$ has fewer columns than
  $M^C$,
  \begin{enumerate}[label=(\arabic{enumi}.\arabic*)]
  \item return $\tilde M = T^{R\dagger}$, $T = M^{CR\dagger} \cdot T^C$,
  \end{enumerate}
\item Else,
  \begin{enumerate}[label=(\arabic{enumi}.\arabic*)]
  \item return $\tilde M = M^C$, $T = T^C$
  \end{enumerate}
\end{enumerate}

\emph{Remark}: During matrix-matrix products $R_{ij} = \sum_k A_{ik}
B_{kj}$, it is helpful and often necessary to set elements of $R$ for
which $|R_{ij}| < \sum_k |A_{ik}| |B_{kj}| \varepsilon_Z$ is true to
zero. This ensures that where we allow cancellation to zero, we do not
introduce additional terms whenever possible.

Step \ref{alg:dln:relaxed} removes the requirement that we cannot
allow cancellation to zero. Step \ref{alg:dln:inner:start} usually
halves the number of rows of $M$, as there are often many zero rows or
rows parallel to previous ones, making the subsequent QR
decompositions both faster and more accurate. Step
\ref{alg:dln:permutation} sorts columns such that those with few
non-zero entries are considered first while attempting to keep an
upper-triangular form. The former helps to find optimal non-cancelling
linear superpositions, while the latter attempts to restore the
usually-preferred triangular form whenever possible. Steps
\ref{alg:dln:reset} and \ref{alg:dln:reset2} reduce numerical errors
by reverting to the input matrix if no improvements have been
found. Finally, steps \ref{alg:dln:rows} and \ref{alg:dln:rows2} often
help to break cyclic dependencies and achieve optimal compression.

\section{Numerical Threshold Values}
The relevant three threshold values are, with the machine precision
$n_\varepsilon \approx 10^{-16}$:
\begin{itemize}
\item $\varepsilon_D$: During delinearisation, a new column has to be
  equal to the original one to within this value, relative to operator
  norms. In practice, we found $\sqrt{n_\varepsilon} \approx 10^{-8}$ to be a suitable value, as
  columns are usually either completely dependent (with very small
  error) or differ substantially (with very large error). Too small a
  threshold will lead to failure to optimise in some cases, as
  numerical noise may become relatively large during a long
  calculation.
\item $\varepsilon_Z$: The delinearisation method is able to work with
  operators of very different orders of magnitude in the same MPO. In
  turn, this means that small terms are not automatically discarded as
  with SVD. This implies that during the various matrix-matrix
  products encountered during MPO compression, special care has to be
  taken to avoid introducing artifical small terms. In practice, we
  found $10^5 n_\varepsilon \approx 10^{-11}$ to work.
\item $\varepsilon_t$: This threshold serves to avoid small
  coefficients in the transfer matrix, which would lead to valid small
  coefficients in the next tensor. For most sensible operators,
  coefficients should be of order one and if this is not possible, it
  may well be desired to keep the components separate rather than
  conflating them into a single column. Our implementation uses a
  value of $10^{-5}$ here.
\end{itemize}

\end{document}